\documentclass[apj]{emulateapj}

\newcommand{\twom}{2M1207b}

\newcommand{\etal}{et al}

\newcommand{\pq}{$P_{q}$}
\newcommand{\peq}{$P_{eq}$}

\newcommand{\teff}{$T_{\rm eff}$}

\newcommand{\kzz}{$K_{\rm zz}$}
\newcommand{\logg}{$\log(g)$}
\newcommand{\rjup}{$R_{\rm Jup}$}
\newcommand{\mjup}{$M_{\rm Jup}$}

\newcommand{\taumix}{$\tau_{\rm mix}$}
\newcommand{\tauchem}{$\tau_{\rm chem}$}
\newcommand{\hp}{$H_{\rm P}$}
\newcommand{\vc}{$V_{\rm c}$}

\makeatletter

\newcommand{\Rmnum}[1]{\expandafter\@slowromancap\romannumeral #1@}
\makeatother

\shorttitle{The Atmosphere of 2M1207\MakeLowercase{b}}
\shortauthors{Barman et al.}

\begin{document}
\bibliographystyle{apj}

\title{The Young Planet-mass Object 2M1207\MakeLowercase{b}:\\ 
           a cool, cloudy, and methane-poor atmosphere}

 \author{Travis S. Barman\altaffilmark{1}, Bruce Macintosh\altaffilmark{2}, Quinn M. Konopacky\altaffilmark{2}, Christian Marois\altaffilmark{3}}
 \altaffiltext{1}{Lowell Observatory, 1400 W. Mars Hill Rd., Flagstaff, AZ 86001 Email: {\tt barman@lowell.edu}}
 \altaffiltext{2}{Lawrence Livermore National Laboratory, 7000 East Avenue, Livermore, CA 94550, USA}
 \altaffiltext{3}{National Research Council Canada, Herzberg Institute of Astrophysics, 5071 West Saanich Road, Victoria, BC V9E 2E7, Canada}

\begin{abstract}
The properties of \twom, a young ($\sim 8$ Myr) planet-mass companion, have
lacked a satisfactory explanation for some time.  The combination of low
luminosity, red near-IR colors, and L-type near-IR spectrum (previously
consistent with \teff\ $\sim$ 1600 K) implies an abnormally small radius.
Early explanations for the apparent underluminosity of \twom\ invoked an
edge-on disk or the remnant of a recent protoplanetary collision.  The
discovery of a second planet-mass object (HR8799b) with similar luminosity and
colors as \twom\ indicates that a third explanation, one of a purely
atmospheric nature, is more likely.  By including clouds, non-equilibrium
chemistry, and low-gravity, an atmosphere with effective temperature consistent
with evolution cooling-track predictions is revealed.  Consequently
\twom, and others like it, requires no new physics to explain nor do they belong
to a new class of objects. Instead they most likely represent the natural
extension of cloudy substellar atmospheres down to low \teff\ and \logg.  If this
{\em atmosphere only} explanation for \twom\ is correct, then very young
planet-mass objects with near-IR spectra similar to field T dwarfs may be
rare.
\end{abstract}


\section{Introduction}

\twom\ is a planet-mass companion orbiting the young brown dwarf 2M1207A at a
projected separation of $\sim 46$ AU \cite[]{Chauvin2004, Chauvin2005}.  As a
member of the $\sim 8$ Myr old TW Hydra association \cite[]{Gizis2002}, \twom\
is one of the youngest companions of its mass ($\sim$ 2 -- 5 \mjup) ever
imaged.  Regardless of how the system formed, \twom\ provides a unique look
into the atmospheric properties of giant planets and brown dwarfs in the very
early stages of their evolution.

Soon after \twom\ was discovered, it was recognized that its low luminosity and
red near-IR colors were seemingly inconsistent.  With m$_K$ = 16.93 $\pm 0.11$
\cite[]{Chauvin2004}, a $K$-band bolometric correction of 3.25 $\pm 0.14$
\cite[]{Mamajek2005}, and a distance of 52.4 $\pm 1.1$ pc
\cite[]{Ducourant2008}, \twom\ has a luminosity of $\log L/L_\odot = -4.73 \pm
0.12$.  With this luminosity and an age of 8$^{+4}_{-3}$ Myr
\cite[]{Chauvin2004}, brown dwarf cooling tracks \cite[]{Baraffe2003}
predict\footnote{The ranges in predicted values come from the luminosity and
age uncertainties.} \teff\ = 936 -- 1090 K, \logg\ = 3.5 -- 3.8 (cgs units),
radius = 1.34 -- 1.43 \rjup, and a mass of 2.3 -- 4.8 \mjup.  The red near-IR
colors (e.g., $H-K = 1.16 \pm 0.24$; Chauvin et al.  2004\nocite{Chauvin2004})
and near-IR spectra of \twom, on the other hand, are indicative of a mid to
late L spectral type and effective temperature of $\sim 1600$K
\cite[]{Mohanty2007, Patience2010}.  Such a combination of high effective
temperature and low luminosity would require a radius of $\sim 0.7 $\rjup,
about a factor of two below evolution model predictions.  Under the assumption
that the higher \teff\ is correct, two possible explanations for the apparent
underluminosity of \twom\ have been put forward.  An edge-on disk, albeit with
unusual characteristics, could provide the necessary 2.5 mags of gray
extinction to accommodate a larger and more reasonable radius
\cite[]{Mohanty2007}.  A more provocative suggestion requires \twom\ to be the
hot afterglow of a recent protoplanetary collision allowing it to be
simultaneously hot and small \cite[]{Mamajek2007}.  

Both of these early explanations for the observed properties of \twom\ hinge
upon the effective temperature being as high as 1600 K.  Earlier estimates of
\teff\ were likely lead astray by the lack of color-\teff\ relationships
properly calibrated for young, planet-mass objects and the lack of model
atmospheres spanning a broad enough range of cloud and chemical parameters to
encompass objects like \twom.  The discovery of HR8799b \cite[]{Marois2008}, a
second planet-mass object with similar near-IR colors and luminosity as \twom,
motivates this new model atmosphere study of \twom, as it is highly unlikely
that either the edge-on disk or recent-collision explanation applies to both
objects.

In this Letter, an atmosphere-only explanation for the observed properties of
\twom\ is presented.  A combination of clouds of modest thickness and
non-equilibrium CO/CH$_4$ ratio is shown to simultaneously reproduce both the
observed photometric and spectroscopic properties of \twom, with bulk
properties consistent with evolution model predictions.  Such an explanation
was touched upon in several recent papers \cite[]{Currie2011, Skemer2011,
Barman2011}, but here the atmosphere of \twom\ is explored in more detail.  

\section{Clouds}

The properties of brown dwarfs and giant planets are known to be influenced by
atmospheric cloud opacity and it is  well established that clouds play a
central role in the transition from spectral types L to T \cite[]{Allard2001}.
Early work on brown dwarf atmospheres approached cloud modeling
phenomenologically, parameterizing the problem with an emphasis on vertical
mixing \cite[]{Ackerman2001}.  Cloudy and cloud-free limits provide useful
insight into the expected spectroscopic and photometric trends but often fail,
unsurprisingly, to match individual brown dwarfs, especially in the L-to-T
transition region \cite[]{Burrows2006}.

\begin{figure}[t]
\plotone{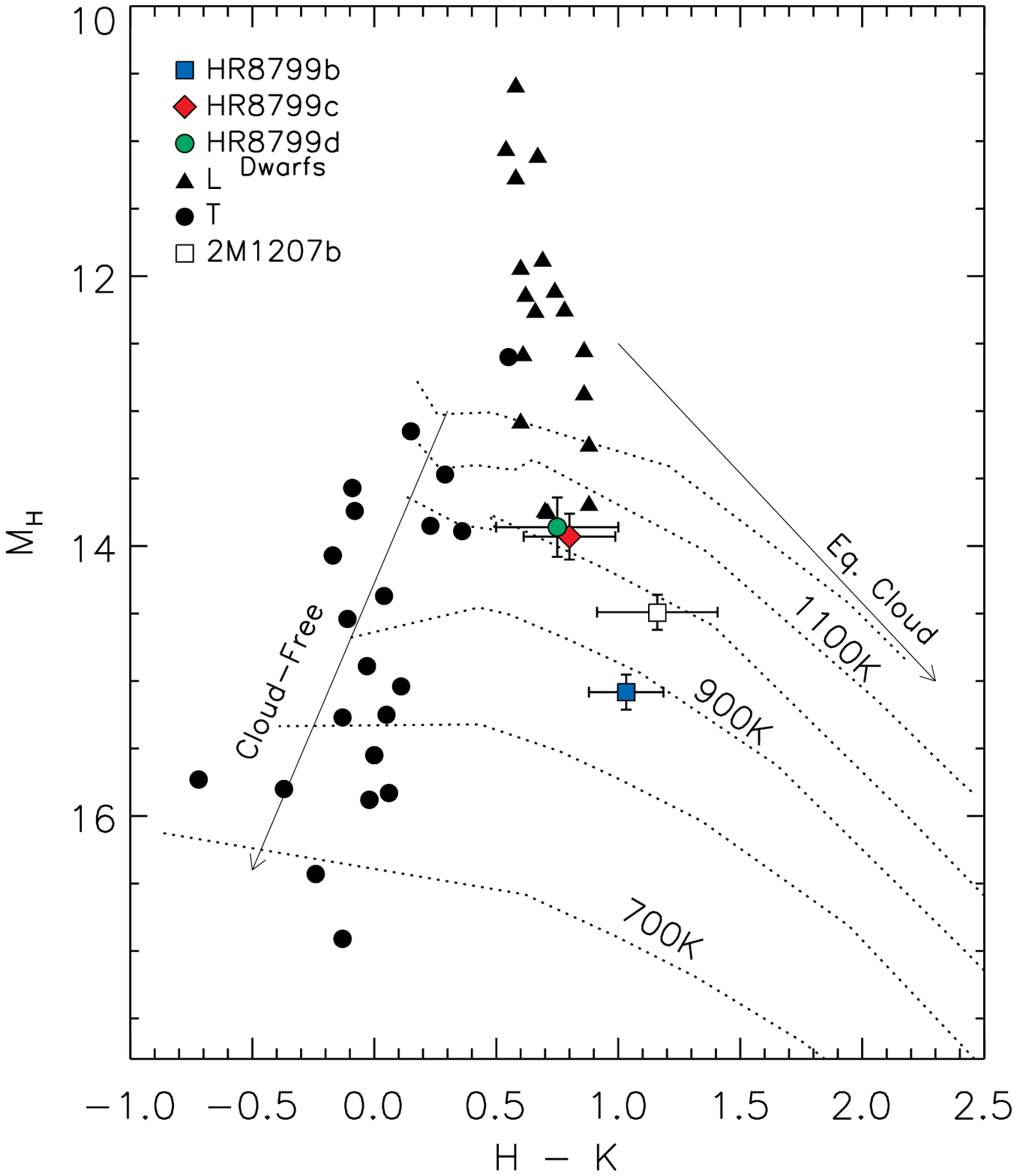}
\caption{Absolute $H$-band magnitude vs. $H$-$K$ near-IR color-magnitude
diagram for field brown dwarfs \cite[]{Leggett2002, Knapp2004}. The planets
\twom\ and HR8799bcd are indicated by symbols with 1-$\sigma$ error bars (see
legend).  Dotted lines show color-magnitude tracks for chemical equilibrium
models with radius = 1.4 \rjup, $\log(g) = 3.5$, mean particle size equal to 5
$\mu$m and T$_{\rm eff}$ equal to 700 -- 1200 K in steps of 100 K, from bottom to
top.  Cloud thickness increases from left to right. The arrows indicate the
approximate locations of cloud free (left) and extremely cloudy (right) models,
with arrow-direction pointing toward decreasing \teff.   The ``transition"
region, between cloudy and cloud-free atmospheres covers a broad range in the
color-magnitude diagram, beyond what is currently occupied by field brown dwarfs.
\label{fig1}}
\end{figure}

Figure \ref{fig1} compares field brown dwarfs to \twom\ and the HR8799 planets
in a color-magnitude diagram (CMD).  \twom\ is located between the cloudy and
cloud-free limits and, consequently, one should not expect either of these
limiting cases to be appropriate when modeling its photometric and
spectroscopic properties.  Figure \ref{fig1} shows the path brown dwarfs or
giant planets of various effective temperatures would follow in a near-IR CMD
if the vertical thickness of clouds is allowed to continuously increase from
zero (cloud-free) to well above the photosphere (pure equilibrium clouds).  The
radius for these tracks, 1.4 \rjup, was specifically chosen to match the radius
predicted for \twom\ as discussed above; however, the paths traced by these
tracks are otherwise independent of evolutionary models.  \twom\ is intersected
by the \teff = 1000 K cloud-track, which is the expected value from evolution
models, demonstrating that low-temperature cloudy atmospheres can achieve very
red near-IR colors, even with clouds significantly thinner than the extreme
limits.

\section{Non-local Equilibrium Chemistry}

If the methane-rich atmospheres of mid to late T dwarfs were in a pure chemical
equilibrium state, CO mole fractions would be too small to have a major
impact on their spectra.  However, CO has been detected in many T dwarfs,
suggesting that their atmospheres are out of equilibrium \cite[]{Noll1997,
Saumon2000, Saumon2006, Geballe2009}.  The most likely mechanism for CO
enhancement is vertical mixing from deep layers, where the temperatures and
pressures are higher and CO is in ready supply.  At photospheric depths and
above, the chemical timescale (\tauchem) to reestablish an equilibrium
CO/CH$_4$ ratio becomes far greater than the mixing timescale (\taumix),
thereby allowing larger CO mole fractions to exist at otherwise
methane-dominated pressures.  Through the same mixing process, N$_2$/NH$_3$ can
also depart from local chemical equilibrium (LCE; Saumon et al. 2006;
Hubeny \& Burrows 2007) \nocite{Saumon2006, Hubeny2007}.  The standard non-LCE
model quenches the CO and CH$_4$ mole fractions at the atmospheric pressure
(\pq) where \taumix = \tauchem, with \taumix\ computed following
\cite{Smith1998}.  Below the quenching depth ($P >$ \pq) the mole fractions for
CO and CH$_4$ are in chemical equilibrium while above ($P \le P_{q}$) they are
set to the values at \pq.

The mole fractions for CO and CH$_4$, N(CO) and N(CH$_4$), at and above \pq\
are very sensitive to the underlying temperature-pressure (T-P)
profile and, thus, are sensitive to gravity, cloud opacity, and metallicity
\cite[]{Hubeny2007, Fortney2008, Barman2011}.  Certain combinations of low
gravity and clouds can result in \pq\ {\em below} the CO/CH$_4$
equilibrium chemistry crossing point (\peq).  When \pq\ is sufficiently deep
and \pq\ $>$ \peq, N(CO) can be quenched at its maximum value while N(CH$_4$)
is quenched near its minimum.  When this situation occurs, the non-LCE
CO/CH$_4$ ratio becomes fairly insensitive to the mixing timescale in the
radiative zone (determined by the adopted coefficient of eddy diffusion, \kzz).
This situation is similar to the N$_2$/NH$_3$ chemistry where the NH$_3$ mole
fractions can also be nearly independent of \kzz, even in high-gravity
cloud-free atmospheres \cite[]{Saumon2006, Hubeny2007}. 

This atmospheric behavior is highly significant to \twom\ as it allows for 
a photosphere with much higher N(CO) and much lower N(CH$_4$) mole fractions at the
low \teff\ predicted by evolution models.  Previous studies, focused primarily
on field brown dwarfs, present far less severe situations with non-LCE only
altering spectra at $\lambda > 4$ $\mu$m where the strongest absorption
bands of CO and NH$_3$ occur \cite[]{Hubeny2007}.  At lower surface gravities,
however, strong non-LCE effects have been shown to extend well into the near-IR
\cite[]{Fortney2008, Barman2011}. 

\section{Model Comparisons}

\twom\ has been observed extensively from the ground and space, with
photometric coverage between 0.9 and $\sim$ 9 \micron\ \cite[]{Chauvin2004,
Song2006, Mohanty2007, Skemer2011}.  High signal-to-noise $J$, $H$, and $K$
spectra are also available \cite[]{Patience2010}.  With the goal of finding a
model atmosphere that agrees with these observations and that has \teff\ and \logg\ in the
range predicted by evolution models, a sequence of atmosphere models was
computed covering \teff\ = 900 -- 1200 K and $\log(g)$ from 3.0 to 4.5 (cgs
units).   The same intermediate cloud (ICM) and non-LCE prescriptions from
\cite{Barman2011} were used.  Synthetic photometry was generated by convolving
synthetic spectra with filter response curves. $J$, $H$, and $K$ spectra were
produced by convolving the model spectra with a Gaussian filter matching the
observed spectral resolution, then interpolated onto the same wavelength
points.  The best-fit was determined by least-squares minimization.

\begin{figure}[t]
\plotone{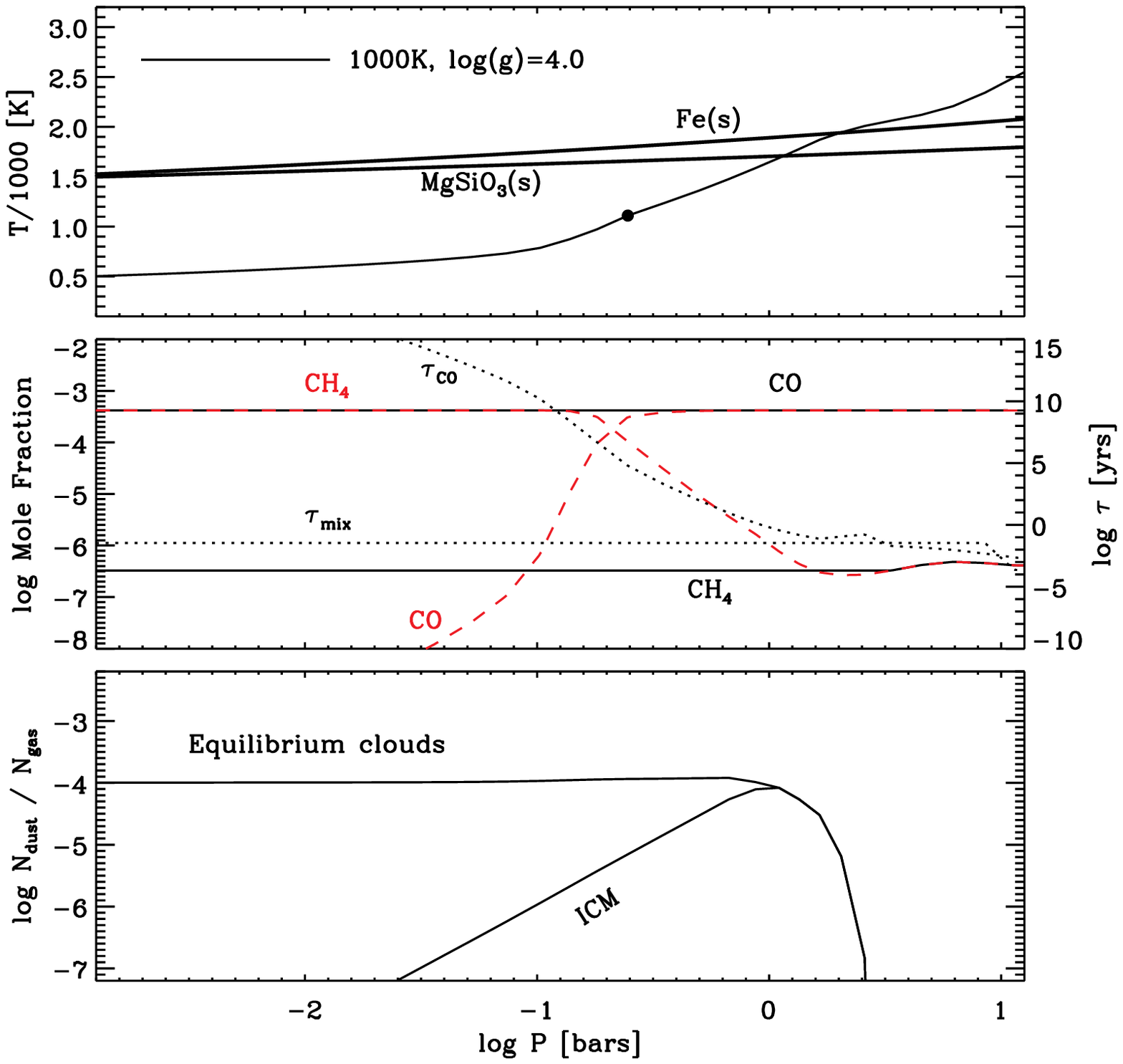}
\caption{
Atmospheric properties for the best fitting model for \twom. {\em Top:}
temperature-pressure structure compared to condensation curves for two abundant
cloud species. {\em Middle:} CO and CH$_4$ mole fractions for equilibrium
(dashed, red) and non-equilibrium (solid, with \kzz\ = 10$^8$ cm$^2$ s$^{-1}$)
chemistry.  Chemical and mixing timescales are also plotted (dotted lines).
{\em Bottom:} dust-to-gas ratio for the intermediate cloud model (ICM) and the
pure equilibrium cloud model.
\label{fig2}}
\end{figure}

Figure \ref{fig2} illustrates the basic structure of the model that best fits the
data (\teff = 1000 K and \logg = 4.0).  The atmospheric cloud, composed mostly
of Fe and Mg-Si grains, has a base at $\sim 3$ bar and extends upward before
dropping off in number-density at $\sim 1$ bar. Despite the rapid drop in
number-density, the cloud extends across the photospheric depths.  Also, 
\pq\ ($\sim 3$ bar) is well below the N(CO)$_{eq}$ = N(CH$_4$)$_{eq}$
point ($P \sim 0.3$ bar), with the CO mole fractions set to the maximum value and
CH$_4$ is close to its minimum value.

While the model cloud and non-LCE properties are determined by free-parameters, they
are likely supported by low gravity and efficient vertical mixing.  In the
convection zone \taumix $\propto$ \hp/\vc, where \vc\ is the convective velocity
and \hp\ is the local pressure scale-height.  With \kzz\ $\propto$
$H_p^2/$\taumix\ $\propto$ \vc\hp, \kzz\ increases with decreasing
gravity in the convection zone as velocity and scale-height increase.  The
radiative-convective boundary also shifts toward the photosphere as surface
gravity decreases, further suggesting that vertical mixing in the radiative
zone near this boundary will also be enhanced in low gravity atmospheres.  In
the radiative zone it is unclear whether the vertical mixing is predominantly
driven by convective overshoot or gravity waves, but the picture emerging from
multi-dimensional hydrodynamical simulations in M dwarfs and brown dwarfs 
indicates that \kzz\ is both depth dependent and easily achieves values $>
10^8$ cm$^2$ s$^{-1}$ \cite[]{Ludwig2006, Freytag2010}. 
 
\begin{figure*}[t]
\epsscale{1.04}
\plotone{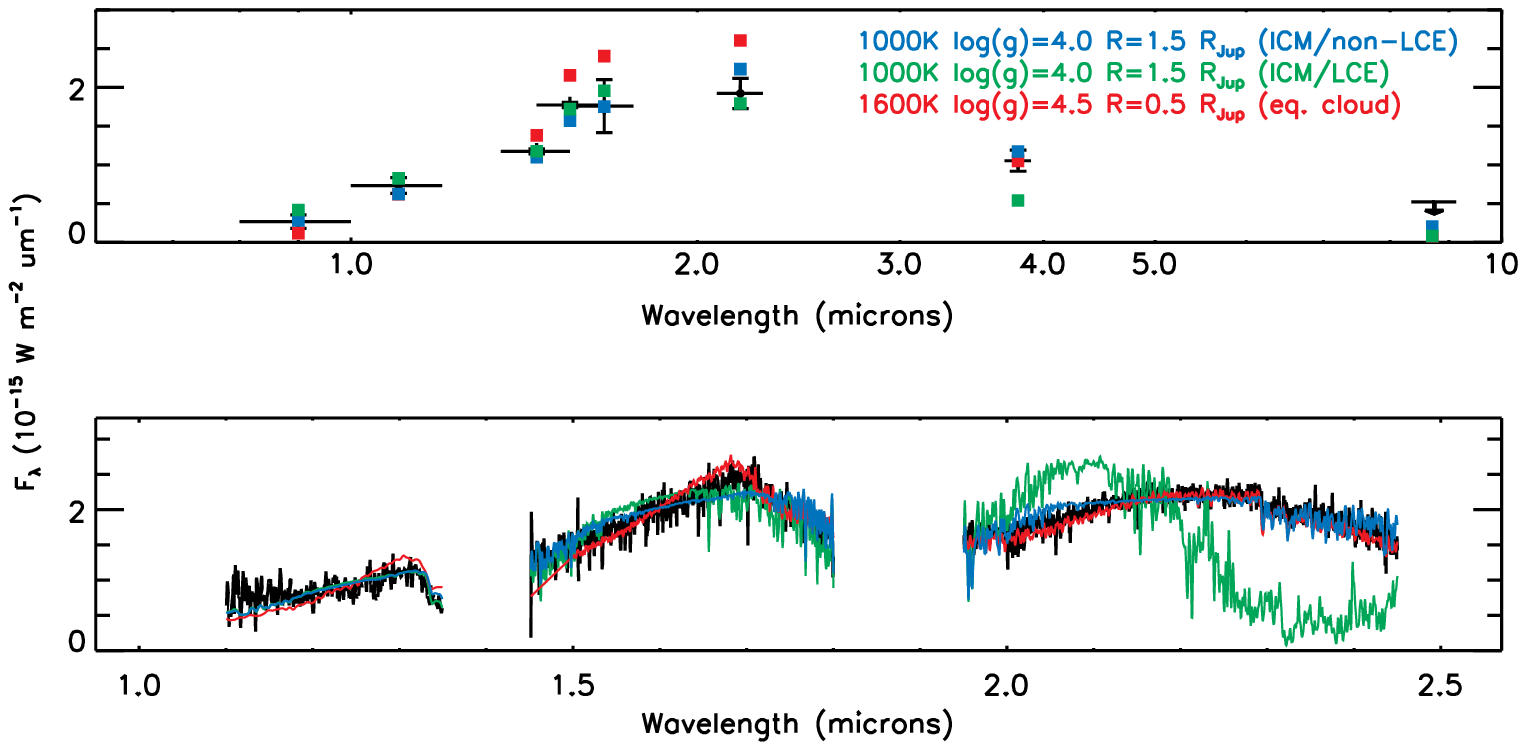}
\caption{
Photometric (top) and spectroscopic (bottom) comparison between \twom\
observations (shown in black, see the text for details) and best-fitting model
(blue).  For comparison, synthetic photometry and spectra from a 1600 K model
(red) with an equilibrium cloud model (aka ``Dusty") along with an LCE model (green) with the
same parameters as the non-LCE model.  Surface gravities, radii, and effective
temperatures are indicated in the legend.  All fluxes have been scaled to 10
pc.
\label{fig3}}
\end{figure*}

Figure \ref{fig3} compares the best-fitting model photometry and spectrum to
the observations.  The model photometry agrees very well with the observations
in nearly ever bandpass, with most bands agreeing at 1$\sigma$.  The model
$J$-band spectrum has about the right slope and nicely reproduces the water
absorption starting at 1.33 $\mu$m.  At $H$ band, the model also has roughly
the correct shape but is slightly too linear across the central wavelengths.
The CO band in the $K$ band is very well reproduced by the model but the
central wavelength region is again too flat.  \cite{Skemer2011} also compare
models \cite[]{Burrows2006, Madu2011} with similar \teff\ and \logg\ to the
same observations; however, these models likely underestimate the non-LCE, as
they do not adequately reproduce the near-IR spectrum, especially the CO band
at 2.3 \micron.  An LCE version of the best-fit non-LCE model is shown in Figure
\ref{fig3} and demonstrates the impact non-LCE has on the near-IR spectrum.

Also shown in Figure \ref{fig3} is the 1600 K ``Dusty" \cite[]{Allard2001}
equilibrium cloud model often selected as the best match to the near-IR spectra
\cite[]{Mohanty2007, Patience2010}.   The 1600 K model photometry does not agree
well with the observations across the full wavelength range.  The 1600 K model
spectrum, however, is very similar to the best-fit 1000 K model across the $K$
band, but only when the near-IR bands are scaled to match individually (to
account for the photometric disagreement for the 1600 K model).  In both $J$ and
$H$ bands, the 1600 K model overpredicts the peak flux and is noticeably more
triangular at $H$ than the observations.  At $K$ band, the 1600 K model provides
only slightly better agreement with the observations than the 1000 K model. 
 
The remaining differences between the 1000 K model and near-IR spectra can be
attributed to an incorrect proportion of dust opacity relative to molecular
opacity.  Given the simplicity of the cloud model used here such discrepancies
are not surprising.  Without a doubt, a more parameter-rich cloud model could
be used to fine tune the comparison, but it is unlikely that such an exercise
will lead to significantly greater insights into the physical properties of the
cloud.  The primary lesson from this comparison is that atmospheric clouds and
chemistry can dramatically alter the spectral shape and potentially lead to
errors in effective temperature as great as 50\%.

The model comparisons to the photometry provide a new estimate for the
bolometric luminosity.  The mean luminosity, found by comparing to pure
equilibrium models, ICM/non-LCE models and black bodies, is $\log L/L_\odot =
-4.68$ with rms of 0.05.  This luminosity is consistent with the earlier value
mentioned above (based on $K$-band bolometric corrections). 

\section{Discussion and Conclusions}

The best fitting \teff, \logg, and radius (1000 K, 10$^4$ cm s$^{-2}$, and 1.5 \rjup)
for \twom\ are consistent with the cooling track predictions discussed above.  This
model demonstrates that including typical cloud thickness and non-LCE are all
that is required to reproduce the current observations of \twom.  Such a model
reminds us that the spectra of brown dwarfs are not strictly a function of
temperature and, at young ages, can deviate significantly from expectations
derived from older field brown dwarfs.  The primary evidence supporting the
edge-on-disk and protoplanetary collision hypotheses, was the previously
deduced 1600 K effective temperature.  A model with this temperature is shown to
inadequately reproduce the available photometry and compares no better to
near-IR spectra than a cooler cloudy model. Also, the disk-model comparison by
\cite{Skemer2011} further weakens the case for an edge-on disk.  Consequently,
no strong evidence remains for the previous disk or collision hypotheses.  This
conclusion is independent of the existence of HR8799b, but is certainly
supported by it.

The primary atmospheric contributors to the L-type appearance of \twom, despite
its low \teff, are clouds extending across the photosphere, thereby reddening
the near-IR colors, and non-equilibrium chemistry, establishing a CO/CH$_4$
ratio that is nearly the reciprocal of what is present in the photospheres of
older field T dwarfs.  The cloud properties, in particular the thickness, are
probably not substantially different from those found in late L dwarfs and the
required \kzz\ ($\sim 10^8$ cm$^2$ s$^{-1}$) is well within the range of
field dwarfs.  \cite{Skemer2011} stress clouds as the primary explanation for
the photometric and spectroscopic properties of \twom.  However, non-LCE plays
an equally important role.  Probably the most important underlying distinction
between \twom\ and field brown dwarfs is low surface gravity, provided by its
youth and low mass.  Without low surface gravity, it is unlikely that clouds or
non-LCE would be sufficient to give \twom\ its current appearance.  Such
objects, therefore, should not be considered members of a new class, but rather
represent the natural extension of substellar atmospheres to low gravity.

Given their similar masses ($\sim 5$\mjup), it is possible that \twom\ and
HR8799b represent two distinct states in the evolution of substellar
atmospheric properties (despite potentially different formation scenarios).
After $\sim 20$ Myr of cooling, perhaps \twom\ will spectroscopically evolve
into something resembling HR8799b and, eventually, into a traditional looking
methane-rich T-dwarf.  The very distinct spectra of the two objects (see Figure
15, \nocite{Barman2011}Barman et al. 2011) would suggest rapid spectral
evolution in the first 50 Myr, post formation.  It is worth noting that the
best-fit models for \twom\ and HR8799b, though similar in atmospheric
parameters, differ in one significant respect -- the radius derived from the
effective temperature and luminosity. Our model here gives a radius of 1.5 \rjup\
for \twom, while even the coldest fit to HR8799b in \cite{Barman2011} gives a
radius of only 1 \rjup.  There are several possible explanations for this.
First, the recovered radius is of course very sensitive to the effective
temperature -- a 100 K increase in \twom\ and a 100 K decrease in HR8799b would
remove the discrepancy, though neither would be a good fit to the spectra.
Second, this may represent a fundamental difference in their internal state due
to different formation scenarios. \twom\ almost certainly did not form through
a core-accretion process as it is highly unlikely that a disk surrounding a
low-mass brown dwarf would have had enough material to accrete a giant planet,
especially at large separations and in a short time.  It is more likely that
\twom\ represents the tail end of binary star formation and, thus, might be
expected to follow a similar cooling evolution as brown dwarfs. The formation
of the HR8799 planets is less clear but could potentially have involved an
accretion period. The ``cold start" accretion models of \cite{Marley2007} predict
significantly smaller radii for a given age and mass. Although the temperature
and luminosity of HR8799b are too high for the extreme cold start models, the
smaller radius may be pointing toward a formation process that involved at
least some loss of entropy.

Finally, one can speculate on the implications that \twom\ and HR8799b might
have on the spectral properties for the broader young brown dwarf and planet
population.  If one adopts, $\sim$ 1500 K as the upper \teff\ limit for
T-dwarfs, then all T dwarfs younger than $\sim 100$ Myr should be in the planet
mass regime ($\lesssim 13$ \mjup) and should have very low gravity ($\log(g)
\lesssim 4.5$).  However, if \twom\ and HR8799b are representative of cool, low
gravity, substellar atmospheres, then non-LCE (and possibly clouds) will
diminish the strength of CH$_4$ absorption across the $H$ and $K$ bands, making
very young methane dwarfs rare.  This prediction, however, is at odds with the
tentative discoveries of $\sim 1$ Myr-old T dwarfs \cite[]{Zap2002,
Burgess2009, Marsh2010}.  While at least one of these discoveries has been
drawn into question \cite[]{Burgasser2004}, if others with strong near-IR
CH$_4$ absorption are confirmed, then it must be explained why substellar
objects of similar age and gravity have atmospheres with wildly different cloud
and non-LCE properties.

\vspace{-0.5cm}
\acknowledgements
We thank the anonymous referee for their review.  This Letter benefited from
many useful discussions with Brad Hansen, Mark Marley, and Didier Saumon.  This
research was supported by NASA through Origins grants to Lowell Observatory and
LLNL along with support from the HST GO program.  Support was also provided by
the NASA High-End Computing (HEC) Program through the NASA Advanced
Supercomputing (NAS) Division at Ames Research Center.  Portions of this work
were performed under the auspices of the U.S. Department of Energy by Lawrence
Livermore National Laboratory under Contract DE-AC52-07NA27344
(LLNL-JRNL-485291).

\end{document}